\documentclass[prd,reprint,showpacs,showkeys]{revtex4}
\usepackage{amsfonts}
\usepackage{amssymb}
\usepackage{amsmath}
\usepackage{graphicx}
\usepackage[font={footnotesize,it}]{caption}

\begin{document}

\title{Classical and quantum analysis of an Einstein-Scalar solution in $2+1$
dimensions}
\author{O. Gurtug}
\email{ozaygurtug@maltepe.edu.tr}
\affiliation{T.C. Maltepe University, Faculty of Engineering and Natural Sciences,
Istanbul, Turkey}
\altaffiliation{Department of Physics, Eastern Mediterranean University, G. Magusa, north
Cyprus, Mersin 10, Turkey. }
\author{S. Habib Mazharimousavi}
\email{habib.mazhari@emu.edu.tr}
\author{M. Halilsoy}
\email{mustafa.halilsoy@emu.edu.tr}
\affiliation{Department of Physics, Eastern Mediterranean University, G. Magusa, north
Cyprus, Mersin 10, Turkey. }

\begin{abstract}
The classical and quantum properties of a new solution obtained in $2+1$%
-dimensional gravity coupled with a real scalar field is analyzed in detail.
The considered new solution is a one-parameter generalization of a
previously known solution. We investigate the solution classically by
calculating exact null and null circular geodesics which correspond to a
typical particle probe of the correspondsing black hole spacetime. The
solution admitting naked singularity is investigated within the context of
quantum mechanics. The timelike naked singularity is probed with spin-zero
and spin-half quantum particles. We show that the curvature singularity in
the new solution is stronger and hence in general the spacetime remains
quantum singular with respect to a quantum particle probe.
\end{abstract}

\maketitle

\section{Introduction}

One of the most important motivation to study general relativity in lower
dimensions, in particular $2+1$-dimension, is to understand and explore
physics in a rather simpler context. Among the others, the well known $2+1$%
-dimensional study in the literature is the solution obtained by Ba\~{n}%
ados, Teitelboim and Zanelli (BTZ) \cite{1}, which describes black hole
solution with a negative cosmological constant. The general properties of
the $2+1$-dimensional gravity, particularly the BTZ solution was considered
in detail by Carlip in \cite{2}. In this seminal paper, Carlip addressed
also the possible relations to quantum gravity via the "microscopic" physics
of quantum gravitational states. No doubt that, this work has unveiled the
underlying rich structure in $2+1$-dimensional gravity and has motivated
many researchers to work along this line.

In the last two decades, $2+1$-dimensional gravity sourced by different
matter fields has attracted considerable interest. Einstein-Maxwell
extension has been considered both in linear \cite{3} and nonlinear \cite%
{4,5,6} electrodynamics. Einstein-scalar and Einstein-Maxwell-scalar
extensions are considered in \cite{7,8} and \cite{9}, respectively. Scalar
field extension in the form of multiplets is also considered in \cite{10}.
In these studies, the solutions admitting black holes are investigated in
details in terms of thermodynamic properties. However, solutions admitting
naked singularities are studied rarely.

The aim of this paper is to study the classical and the quantum properties
of the recently obtained solution that incorporates with a self-interacting
real scalar field \cite{10}. This solution is a one parameter generalization
of the solution obtained by Schmidt and Singleton (SS) in \cite{8}. The SS
solution does not have a horizon and the curvature singularity at $r=0$ is a
typical naked singularity. Another notable feature of the SS solution is
that the spatial part of the metric is flat, whereas the temporal part is
equivalent to the asymptotic form of AdS spacetime. In contrast to the SS
solution, the recently found metric in \cite{10}, can possesses black hole
or naked singularity depending on the parameters. Furthermore, the spatial
part of the metric given in \cite{10} is not flat. Hence, the inclusion of a
parameter in the new solution has a drastic effect on the topology of the
resulting spacetime.

In this paper; first as a classical analysis, the physical properties of the
spacetime geometry of a self-gravitating real scalar field admitting black
hole solution will be investigated by considering the geodesics equations.
The null and timelike geodesics are obtained exactly by integrating the
geodesic equations.

Secondly; as a quantum analysis, the occurrence of naked singularity that
the solution given in \cite{10} admits will be investigated. Solutions
admitting naked singularities are always undervalued. Hence, our knowledge
about their structure / nature is limited. This limited information is in
fact natural, because the scale where this singularity forms is smaller than
the Planck scale. In this microscopic regime, one would not expect from the
classical methods to provide reliable insight. The general belief in these
regimes where the curvature of the spacetime becomes enormous is that to
adopt the laws of quantum gravity. Consequently, the appropriate method for
analyzing the singularities must be found within the framework of quantum
theory of gravity. Unfortunately, there is no consistent quantum theory of
gravity to be used in singularity analysis yet. In the light of this fact,
any method that incorporates with quantum fields / particles for the
resolution or persistence of curvature singularities would be considered as
a step in the right direction. In the literature, there are alternative
methods for this purpose. Loop quantum gravity \cite{11} and string theory 
\cite{12,13} are the two important fields in the resolution of
singularities. In this paper, an alternative method will be used,so that the
motion of quantum fields / particles will be considered in a classical
background. In the analysis of the naked singularity, $spin$ $0$ (scalar)
and $spin$ $1/2$ (fermionic) wave packets obeying the Klein-Gordon and Dirac
equations will be used in probing the singularity. Instead of a point
particle probe which leads to the notion of \textit{geodesics incompleteness}%
, the quantum particle probe will be used which leads to the notion of 
\textit{quantum singularity}.

Organization of the paper is as follows. In Section II we briefly review the
new Einstein-scalar solution in $2+1-$dimensions. Geodesics of a classical
particle in the metric admitting black hole is considered in Section III.
Quantum probes of spin$-0$ and spin$-\frac{1}{2}$ are considered for the
timelike naked singular version of the solution in Section IV. The paper is
completed with Conclusion and Discussion in Section V.

\section{The solution in brief}

Recently, a new exact radial solution with a self-interacting real scalar
field in $2+1$-dimensional gravity has been introduced in \cite{10}. This
new solution in fact constitutes a one-parameter generalization of the
solution obtained by Schmidt and Singleton (SS) in \cite{8}. The striking
property of the solution presented in \cite{8} is that the spatial part of
the metric was flat. However, in the new solution \cite{10}, the spatial
part is not flat, i.e. the metric tensor component $g_{rr}\neq 1$. In
comparison with the solution given by SS, the new solution has a richer
structure. One of the distinctive richness is the existence of black holes.

The action and the line element ansatz in Einstein-scalar theory in $2+1$%
-dimensions are given by 
\begin{equation}
S=\int \sqrt{-g}\left( R-\frac{1}{2}\left( \nabla \phi \right) ^{2}-V\left(
\phi \right) \right) d^{3}x,
\end{equation}%
\begin{equation}
ds^{2}=-A(r)dt^{2}+\frac{dr^{2}}{A(r)}+H(r)^{2}d\theta ^{2},
\end{equation}%
respectively in which 
\begin{equation}
\phi (r)=\frac{2\alpha }{2+\alpha ^{2}}\ln r,
\end{equation}%
and 
\begin{equation}
V\left( \phi \right) =V_{0}e^{-\alpha \phi }=V_{0}\left( \frac{1}{r}\right)
^{\frac{2\alpha ^{2}}{2+\alpha ^{2}}}.
\end{equation}%
Furthermore in \cite{10} it was shown that the solution of the field
equations gives 
\begin{equation}
A\left( r\right) =\frac{V_{0}\left( 2+\alpha ^{2}\right) ^{2}}{2\left(
\alpha ^{2}-4\right) }r^{\frac{4}{2+\alpha ^{2}}}+Cr^{\frac{\alpha ^{2}}{%
2+\alpha ^{2}}}
\end{equation}%
and 
\begin{equation}
H\left( r\right) =r^{\frac{4}{2+\alpha ^{2}}}.
\end{equation}%
In these solutions $\alpha \neq \pm 2$ and $V_{0}$ are two real parameters
and $C$ is an integration constant.

Finally, the line element in terms of the parameters can be written as%
\begin{equation}
ds^{2}=-\left( \frac{V_{0}\left( \alpha ^{2}+2\right) ^{2}}{2\left( \alpha
^{2}-4\right) }r^{\frac{4}{2+\alpha ^{2}}}+Cr^{\frac{\alpha ^{2}}{2+\alpha
^{2}}}\right) dt^{2}+\frac{dr^{2}}{\left( \frac{V_{0}\left( \alpha
^{2}+2\right) ^{2}}{2\left( \alpha ^{2}-4\right) }r^{\frac{4}{2+\alpha ^{2}}%
}+Cr^{\frac{\alpha ^{2}}{2+\alpha ^{2}}}\right) }+r^{\frac{4}{2+\alpha ^{2}}%
}d\theta ^{2}
\end{equation}%
which is in general highly dependent on the three parameters $\alpha ,$ $C$
and $V_{0}.$ Depending on the choice of these parameters, the general
solution admits black holes or naked singularities. For instance, the black
hole solution is possible if one sets $\alpha =\sqrt{2}$ and $C<0$ with $%
V_{0}<0$ where a quasilocal mass evaluated from the Brown-York \cite{14}
formalism is given by%
\begin{equation}
M_{QL}=\frac{\left\vert C\right\vert }{8}.
\end{equation}%
Note that if we take $\alpha =\sqrt{2}$ we obtain the following line element%
\begin{equation}
ds^{2}=-\left( C\sqrt{r}-4V_{0}r\right) dt^{2}+\frac{dr^{2}}{\left( C\sqrt{r}%
-4V_{0}r\right) }+rd\theta ^{2}.
\end{equation}%
After redefining the radial coordinate as $r=\rho ^{2},$ the metric (9)
becomes%
\begin{equation}
ds^{2}=-\rho \left( C-4V_{0}\rho \right) dt^{2}+\frac{4\rho }{\left(
C-4V_{0}\rho \right) }d\rho ^{2}+\rho ^{2}d\theta ^{2}.
\end{equation}%
Before going further let us justify our choice of $\alpha $ by the fact that 
$\frac{\partial \phi }{\partial \alpha }=0,$ and $\frac{\partial ^{2}\phi }{%
\partial \alpha ^{2}}<0,$ at $\alpha ^{2}=2,$ which provides the maximum
effect for this specific $\alpha .$ The resulting metric given in equation
(10), can be interpreted as a one-parameter generalization (namely, with $C$%
) of the solution obtained by SS. One can easily recover the solution in SS,
if $C=0,$ $V_{0}=-1$ \cite{8}. Obviously, the resulting metric (10) admits
naked singularities if $C>0$, $V_{0}<0$ or $C<0$, $V_{0}>0.$ One of the
scope of this paper is to investigate the nature of this naked singularity
in view of quantum mechanics.

\section{Geodesic motion of a classical particle}

As we mentioned in previous section, the general solution for $\alpha ^{2}<4$
admits black hole with a general line element given by \cite{AMH}%
\begin{equation}
ds^{2}=-f\left( r\right) dt^{2}+\frac{r^{\alpha ^{2}}}{f\left( r\right) }%
dr^{2}+r^{2}d\theta ^{2}
\end{equation}%
in which 
\begin{equation}
f\left( r\right) =\frac{4}{4-\alpha ^{2}}\frac{r^{2}}{\ell ^{2}}\left(
1-\left( \frac{r_{h}}{r}\right) ^{\frac{4-\alpha ^{2}}{2}}\right)
\end{equation}%
with 
\begin{equation}
r_{h}=\left( \frac{\ell ^{2}\left( 4-\alpha ^{2}\right) M_{BTZ}}{4}\right) ^{%
\frac{2}{4-\alpha ^{2}}}.
\end{equation}%
Herein $M_{BTZ}$ guarantees that in the limit $\alpha \rightarrow 0$ we get
the corresponding BTZ solution with the proper mass \cite{AMH}. In addition
to these, in terms of the parameter used in previous section one finds%
\begin{equation}
\ell ^{2}=\frac{8}{\left\vert V_{0}\right\vert \left( \alpha ^{2}+2\right)
^{2}}
\end{equation}%
and%
\begin{equation}
r_{h}=\left( \frac{2\left\vert C\right\vert \left( 4-\alpha ^{2}\right) }{%
\left\vert V_{0}\right\vert \left( \alpha ^{2}+2\right) ^{2}}\right) ^{\frac{%
2}{4-\alpha ^{2}}}
\end{equation}%
provided both $C$ and $V_{0}$ are negative.

The Lagrangian of a classical test particle with unit mass is given by%
\begin{equation}
L=\frac{1}{2}\left( -f\left( r\right) \dot{t}^{2}+\frac{r^{\alpha ^{2}}}{%
f\left( r\right) }\dot{r}^{2}+r^{2}\dot{\theta}^{2}\right)
\end{equation}%
in which a dot stands for the derivative with respect to the proper time $%
\tau .$ The normalization condition, in other hands, imposes $U^{\mu }U_{\mu
}=\epsilon $ or%
\begin{equation}
-f\left( r\right) \dot{t}^{2}+\frac{r^{\alpha ^{2}}}{f\left( r\right) }\dot{r%
}^{2}+r^{2}\dot{\theta}^{2}=\epsilon
\end{equation}%
in which $U^{\mu }$ is the four-velocity and $\epsilon =0,\pm 1$ for null,
timelike / spacelike geodesics. The Euler Lagrange equation can be found by
introducing%
\begin{equation}
P_{0}=-E=\frac{\partial L}{\partial \dot{t}}=-f\dot{t}
\end{equation}%
and%
\begin{equation}
P_{2}=J=\frac{\partial L}{\partial \dot{\theta}}=r^{2}\dot{\theta}
\end{equation}%
in which $E$ is the energy of the particle and $J$ the angular momentum.
Inserting (19) and (18) into (17), we find the following%
\begin{equation}
\dot{r}^{2}=\left( \epsilon +\frac{E^{2}}{f}-\frac{J^{2}}{r^{2}}\right) 
\frac{f}{r^{\alpha ^{2}}}.
\end{equation}%
A combination of (19) and (20) admits 
\begin{equation}
\left( \frac{dr}{d\theta }\right) ^{2}=\left( \frac{\epsilon }{J^{2}}+\frac{%
E^{2}}{J^{2}f}-\frac{1}{r^{2}}\right) \frac{f}{r^{\alpha ^{2}-4}}.
\end{equation}

\subsection{Null geodesics}

Setting $\epsilon =0$ in (21) one finds%
\begin{equation}
\left( \frac{dr}{d\theta }\right) ^{2}=\frac{\xi ^{2}r^{2}-f}{r^{\alpha
^{2}-2}}
\end{equation}%
in which $\xi ^{2}=\frac{E^{2}}{J^{2}}.$ Considering (12) the latter
equation reads%
\begin{equation}
\left( \frac{dr}{d\theta }\right) ^{2}=\left( \xi ^{2}-\frac{4}{\left(
4-\alpha ^{2}\right) \ell ^{2}}\right) r^{4-\alpha ^{2}}+\frac{4}{\left(
4-\alpha ^{2}\right) \ell ^{2}}\left( rr_{h}\right) ^{\frac{4-\alpha ^{2}}{2}%
}.
\end{equation}%
To proceed further we set $\alpha ^{2}=2$ resulting in 
\begin{equation}
\left( \frac{dr}{d\theta }\right) ^{2}=\left( \xi ^{2}-\frac{2}{\ell ^{2}}%
\right) r^{2}+\frac{2}{\ell ^{2}}rr_{h}
\end{equation}%
which upon introducing the new variables $\tilde{r}=\frac{\ell ^{2}}{2r_{h}}%
\left( \xi ^{2}-\frac{2}{\ell ^{2}}\right) r$ and $\tilde{\theta}=\theta 
\sqrt{\xi ^{2}-\frac{2}{\ell ^{2}}}$ this yields%
\begin{equation}
\left( \frac{d\tilde{r}}{d\tilde{\theta}}\right) ^{2}=\tilde{r}^{2}+\tilde{r}%
.
\end{equation}%
This admits a solution given by%
\begin{equation}
\tilde{r}_{\pm }=\frac{e^{\pm \left( \tilde{\theta}+\tilde{C}\right) }}{2}-%
\frac{1}{2}+\frac{e^{\mp \left( \tilde{\theta}+\tilde{C}\right) }}{8}
\end{equation}%
in which $\tilde{C}$ is an integration constant.

Finally we add that the inverse transformation of the variables yield a
complete solution expressed by%
\begin{equation}
r_{\pm }=\frac{e^{\pm \sqrt{\xi ^{2}-\frac{2}{\ell ^{2}}}\left( \theta +\eta
\right) }-1+\frac{1}{4}e^{\mp \sqrt{\xi ^{2}-\frac{2}{\ell ^{2}}}\left(
\theta +\eta \right) }}{\ell ^{2}\left( \xi ^{2}-\frac{2}{\ell ^{2}}\right) }%
r_{h}
\end{equation}%
where $\tilde{C}=\eta \sqrt{\xi ^{2}-\frac{2}{\ell ^{2}}}$ provided $\xi
^{2}-\frac{2}{\ell ^{2}}>0.$

In null geodesics when $\epsilon =0$ the coordinate time $t$ equation of the
position $r$ is given by the combination of Eqs. (20) and (18). Hence, one
obtains%
\begin{equation}
\frac{dr}{dt}=\pm \sqrt{1-\frac{J^{2}f}{E^{2}r^{2}}}\frac{f}{r^{\alpha
^{2}/2}}
\end{equation}%
where for $\alpha ^{2}=2$ and $f$ given by (12) we find%
\begin{equation}
\frac{dr}{dt}=\pm \sqrt{1-\frac{2J^{2}}{E^{2}\ell ^{2}}+\frac{2J^{2}}{%
E^{2}\ell ^{2}}\frac{r_{h}}{r}}\frac{2}{\ell ^{2}}\left( r-r_{h}\right)
\end{equation}%
which upon a change of variable of the form $r=\frac{r_{h}}{u}$ becomes%
\begin{equation}
\frac{du}{dt}=\mp \sqrt{1-\frac{2J^{2}}{E^{2}\ell ^{2}}+\frac{2J^{2}}{%
E^{2}\ell ^{2}}u}\frac{2u}{\ell ^{2}}\left( 1-u\right) .
\end{equation}%
This equation admits an exact solution given by%
\begin{eqnarray}
\mp \frac{t-t_{0}}{\ell ^{2}} &=&\tanh ^{-1}\frac{\sqrt{1-\frac{2J^{2}}{%
E^{2}\ell ^{2}}+\frac{2J^{2}}{E^{2}\ell ^{2}}u}-\sqrt{1-\frac{2J^{2}}{%
E^{2}\ell ^{2}}+\frac{2J^{2}}{E^{2}\ell ^{2}}u_{0}}}{1-\sqrt{1-\frac{2J^{2}}{%
E^{2}\ell ^{2}}+\frac{2J^{2}}{E^{2}\ell ^{2}}u}\sqrt{1-\frac{2J^{2}}{%
E^{2}\ell ^{2}}+\frac{2J^{2}}{E^{2}\ell ^{2}}u_{0}}}+ \\
&&\frac{1}{\sqrt{1-\frac{2J^{2}}{E^{2}\ell ^{2}}}}\tanh ^{-1}\left( \frac{%
\sqrt{1-\frac{2J^{2}}{E^{2}\ell ^{2}}}\left( \sqrt{1-\frac{2J^{2}}{E^{2}\ell
^{2}}+\frac{2J^{2}}{E^{2}\ell ^{2}}u}-\sqrt{1-\frac{2J^{2}}{E^{2}\ell ^{2}}+%
\frac{2J^{2}}{E^{2}\ell ^{2}}u_{0}}\right) }{1-\frac{2J^{2}}{E^{2}\ell ^{2}}-%
\sqrt{1-\frac{2J^{2}}{E^{2}\ell ^{2}}+\frac{2J^{2}}{E^{2}\ell ^{2}}u}\sqrt{1-%
\frac{2J^{2}}{E^{2}\ell ^{2}}+\frac{2J^{2}}{E^{2}\ell ^{2}}u_{0}}}\right)
\end{eqnarray}%
in which $u_{0}=u\left( t=t_{0}\right) .$ As $u=\frac{r_{h}}{r}$ is smaller
than $1$ i.e., $u<1$, so is $u_{0}$ which imply that $1>\frac{2J^{2}}{%
E^{2}\ell ^{2}}.$ At exactly $\frac{2J^{2}}{E^{2}\ell ^{2}}=1$ the solution
is obtained to be%
\begin{equation}
\mp \frac{t-t_{0}}{\ell ^{2}}=-\left( x-x_{0}\right) +\ln \left( \frac{%
\left( x+1\right) \left( x_{0}-1\right) }{\left( x-1\right) \left(
x_{0}+1\right) }\right)
\end{equation}%
in which $x=\left( \frac{r}{r_{h}}\right) ^{2}=\frac{1}{u^{2}}>1$ and $%
x_{0}=x\left( t=t_{0}\right) .$ Note that the negative branch corresponds to 
$\frac{dr}{dt}>0$ and vice versa. From the latter equation we comment that
in the negative branch solution the null particle escapes from the black
hole and in an infinite time interval it reaches to infinity. In contrary,
the positive branch solution approaches to the horizon in an infinite time
interval. These are shown in Fig. 1.

%
%
%
%
%
\begin{figure}[tbp]
\includegraphics[width=120mm,scale=0.7]{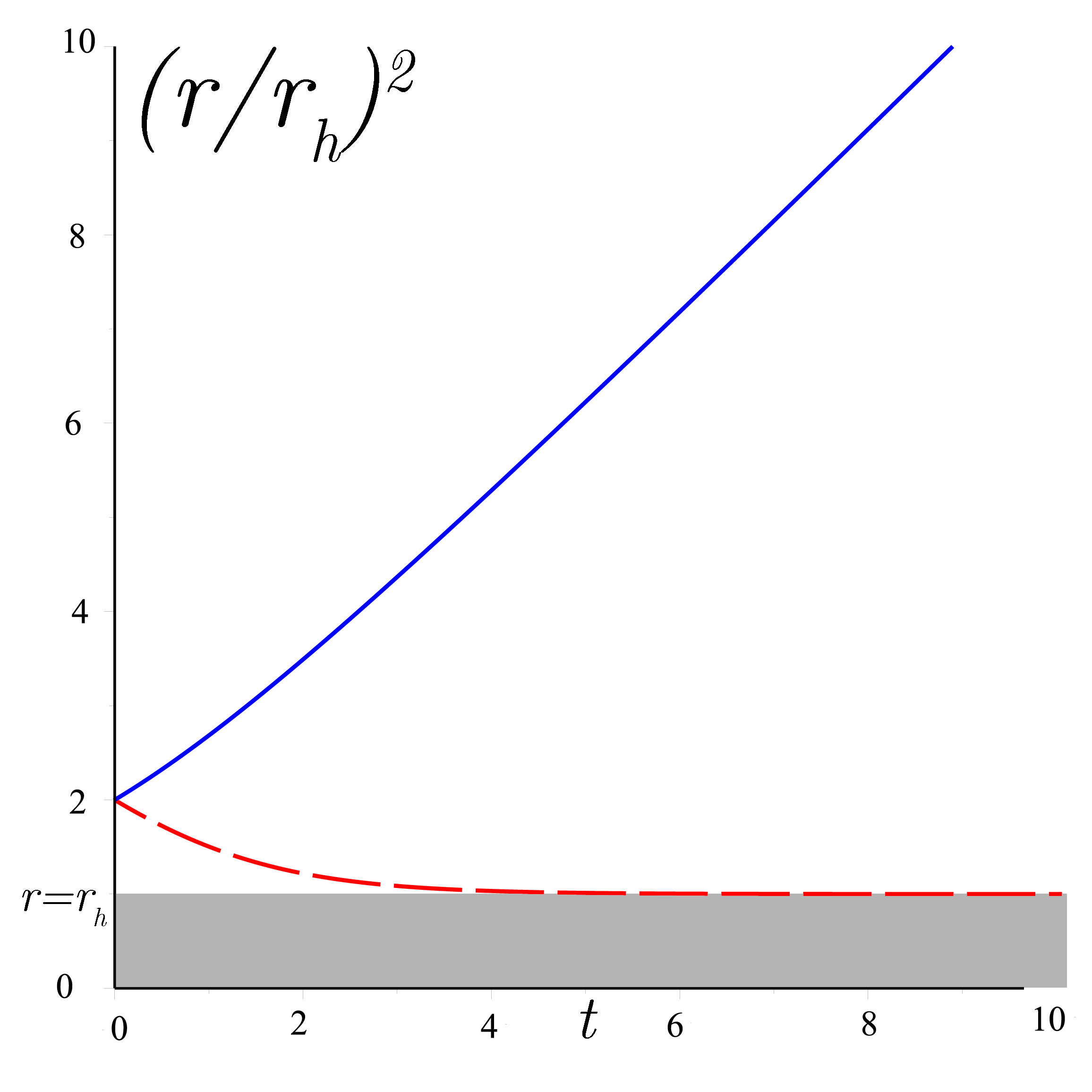}
\caption{$x\left( t\right) =\left( \frac{r}{r_{h}}\right) ^{2}$ versus $%
t/\ell ^{2}$ for positive (Red-Dash) and negative (Blue-Solid) branch
solutions corresponding to the negative and positive initial velocity
respectively and $\frac{2J^{2}}{E^{2}\ell ^{2}}=1$. }
\end{figure}

\subsection{Null circular geodesics}

In circular motion we have $r=r_{c}$ and is constant. This together with
(24) we find%
\begin{equation}
\left( \xi ^{2}-\frac{2}{\ell ^{2}}\right) r_{c}+\frac{2}{\ell ^{2}}r_{h}=0.
\end{equation}%
This equation admits a solution for $r_{c}$ given by%
\begin{equation}
r_{c}=\frac{1}{1-\frac{\ell ^{2}E^{2}}{2J^{2}}}r_{h}
\end{equation}%
which for $1>\frac{\ell ^{2}E^{2}}{2J^{2}}$ the solution is positive and
larger than the radius of event horizon. Therefore for null particle with
given $\frac{E^{2}}{J^{2}}$ there exist only one single possible circular
orbit whose radius is obtained as (35). Looking at the master radial
equation for the null particle with affine parameter $\lambda $ 
\begin{equation}
\left( \frac{dr}{d\lambda }\right) ^{2}=\frac{E^{2}}{r^{2}}-\frac{J^{2}f}{%
r^{4}}
\end{equation}%
and rewriting it as%
\begin{equation}
\left( \frac{dr}{d\lambda }\right) ^{2}=\frac{2J^{2}}{\ell ^{2}}\frac{1}{%
r^{2}}\left( \frac{r_{h}}{r}-\frac{r_{h}}{r_{c}}\right)
\end{equation}%
one can find 
\begin{equation}
\frac{d^{2}r}{d\lambda ^{2}}=\frac{J^{2}}{\ell ^{2}}\frac{2r_{h}}{r^{3}r_{c}}%
\left( 1-\frac{3r_{c}}{2r}\right)
\end{equation}%
which is clearly negative at $r=r_{c},$ indicating that the circular motion
is unstable.

\subsection{Null radial geodesics}

For the radial motion of a null particle we set $J=0$ in (29) which yields%
\begin{equation}
\frac{dr}{dt}=\pm \frac{2}{\ell ^{2}}\left( r-r_{h}\right)
\end{equation}%
whose exact solution is given by%
\begin{equation}
\frac{r}{r_{h}}=1+\left( \frac{r_{0}}{r_{h}}-1\right) e^{\pm \frac{2}{\ell
^{2}}\left( t-t_{0}\right) }
\end{equation}%
in which $r\left( t=t_{0}\right) =r_{0}.$ In the case of outward initial
velocity the radial position of the null particle increases as fast as the
exponential of time and for the inward initial velocity it approaches
asymptotically to the horizon.

\section{Quantum Probe of the Solution Admitting Naked Singularity}

Spacetime singularities constitutes one of the important predictions of the
Einstein's theory of relativity. It describes the location in the fabric of
spacetime in which the physical quantities like tidal forces, gravitational
fields and energy densities becomes unbounded. Or, in other words, it is
defined as the "end point" for the time evolution of the timelike and null
geodesics. \ According to the Penrose's weak cosmic censorship hypothesis,
singularities must be covered by horizon(s), for the deterministic nature of
the theory of general relativity. However, there are some situations such
that the solutions to the Einstein's equations admit singularities that are
not hidden by horizon(s) and becomes visible to the asymptotic observers.
Singularities in this kind are called naked singularities and violates the
cosmic censorship hypothesis of Penrose. A question has been asked: how
could this singularity problem in general relativity be resolved ?.

So far, it seems that the most promising method for the resolution of the
singularities is the quantum theory of gravity. The reason for this is the
scales where these singularities are developed, and hence, classical general
relativity is expected to be replaced by quantum theory of gravity. However,
a complete quantum theory of gravity is not available yet. For this reason,
an alternative method which was developed by Horowitz and Marolf (HM) \cite%
{15} following the work of Wald \cite{16} will be used in this paper. In
order to probe the naked singularity, quantum particles with $spin-0$
(scalar) and $spin-1/2$ (fermion) comply with Klein - Gordon and Dirac
equations will be used, respectively.

The main theme of the HM criterion which can be applied only to static
spacetimes having timelike naked singularities is summarized as follows: The
crucial idea is to split the spatial and time part of the Klein-Gordon
equation and write it in the form of%
\begin{equation}
\frac{\partial ^{2}\psi }{\partial t^{2}}=-A\psi ,
\end{equation}%
where $A$ is the spatial wave operator. Note that, this operator is a
symmetric and positive on the Hilbert space $\mathcal{H}.$ According to the
HM, the singular character of the spacetime with respect to quantum probe is
characterized by investigating whether the spatial part of the operator $A$
has a unique self - adjoint extension (i.e. essentially self - adjoint) in
the entire space or not. If the extension is unique, it is said that the
space is quantum mechanically regular. In order to clarify this point,
consider the Klein-Gordon equation for a free particle that satisfies%
\begin{equation}
i\frac{d\psi }{dt}=\sqrt{A_{E}}\psi ,
\end{equation}%
whose solution is%
\begin{equation}
\psi \left( t\right) =e^{-it\sqrt{A_{E}}}\psi \left( 0\right) ,
\end{equation}%
in which $A_{E}$ denotes the extension of the operator $A$. If $A$ has not a
unique self - adjoint extension, then the future time evolution of the wave
function (42) is ambiguous and HM criterion defines the spacetime as quantum
mechanically singular . The essential self-adjointness of the operator $A$,
can be verified by considering solutions of the equation%
\begin{equation}
A^{\ast }\psi \pm i\psi =0,
\end{equation}%
(with $A^{\ast }$ denoting the adjoint of $A$) and showing that the
solutions of (44), do not belong to Hilbert space $\mathcal{H}$ (we refer to 
\cite{17,20} for detailed mathematical analysis). This will be achieved by
defining the function space on each $t=$constant hypersurface $\Sigma $ as $%
\mathcal{H}=\{R:\left\Vert R\right\Vert <\infty \}$ with the following norm
given for the metric (10) as,%
\begin{equation}
\left\Vert R\right\Vert ^{2}=\int_{0}^{\infty }\sqrt{\frac{g_{\rho \rho
}g_{\theta \theta }}{g_{tt}}}\left\vert R\right\vert ^{2}d\rho
=\int_{0}^{\infty }\frac{\rho }{A(\rho )}\left\vert R\right\vert ^{2}d\rho .
\end{equation}

\subsection{Klein-Gordon Fields}

The timelike naked singularity for the metric (10) will be probed with
massive scalar particles (bosons) satisfying the Klein-Gordon equation%
\begin{equation}
\left( \frac{1}{\sqrt{g}}\partial _{\mu }\left[ \sqrt{g}g^{\mu \nu }\partial
_{\nu }\right] -\tilde{m}^{2}\right) \psi =0,
\end{equation}%
in which $\tilde{m}$ is the mass of the scalar particle. Equation (46) for
the metric (10) is given below by splitting temporal and spatial part as%
\begin{equation}
\frac{\partial ^{2}\psi }{\partial t^{2}}=-\left\{ -\left( \frac{%
C-4V_{0}\rho }{2}\right) ^{2}\frac{\partial ^{2}\psi }{\partial \rho ^{2}}-%
\left[ \frac{\left( C-4V_{0}\rho \right) \left( C-8V_{0}\rho \right) }{4\rho 
}\right] \frac{\partial \psi }{\partial \rho }-\left( \frac{C-4V_{0}\rho }{%
\rho }\right) \frac{\partial ^{2}\psi }{\partial \theta ^{2}}-\tilde{m}%
^{2}\psi \right\} .
\end{equation}%
By writing the Klein-Gordon equation in the form of (41), one can easily
write the spatial part of the wave operator as%
\begin{equation}
A=-\left( \frac{C-4V_{0}\rho }{2}\right) ^{2}\frac{\partial ^{2}}{\partial
\rho ^{2}}-\left[ \frac{\left( C-4V_{0}\rho \right) \left( C_{2}-8V_{0}\rho
\right) }{4\rho }\right] \frac{\partial }{\partial \rho }-\left( \frac{%
C-4V_{0}\rho }{\rho }\right) \frac{\partial ^{2}}{\partial \theta ^{2}}-%
\tilde{m}^{2}
\end{equation}

The problem now is to count the number of extensions of the operator $A.$
This is done by using the concept of deficiency indices discovered by Weyl 
\cite{21} and generalized by von Neumann \cite{22} . If there are no square
integrable ($L^{2}\left( 0,\infty \right) $) solutions (i.e. with deficiency
indicies $n_{+}=n_{-}=0$) in the entire space, the operator $A$ possesses a
unique self-adjoint extension and it is called essentially self-adjoint.
Consequently, the method to find a sufficient condition for the operator $A$
to be essentially self-adjoint is to investigate the solutions satisfying
equation (44)\ that do not belong to the Hilbert space $\mathcal{H}$. Using
separation of variables, $\psi =R(\rho )Y(\theta ),$ equation (44) yields
the following radial equation 
\begin{equation}
\frac{\partial ^{2}R\left( \rho \right) }{\partial \rho ^{2}}+\left[ \frac{%
C-8V_{0}\rho }{\rho \left( C-4V_{0}\rho \right) }\right] \frac{\partial
R(\rho )}{\partial \rho }+\left( \frac{4}{C-4V_{0}\rho }\right) \left[ \frac{%
\tilde{m}^{2}\pm i}{C-4V_{0}\rho }-\frac{c^{2}}{\rho }\right] R(\rho )=0,
\end{equation}%
in which $c$ denotes the separation constant. Because of the complexity in
finding exact analytic solution to Eq. (49), we study the behavior of $%
R\left( \rho \right) $ near $\rho \rightarrow 0$ and $\rho \rightarrow
\infty .$

\subsubsection{\textbf{The case when} $\protect\rho \rightarrow 0$}

The behaviour of equation (49) when $\rho \rightarrow 0$ is given by

\begin{equation}
\frac{\partial ^{2}R\left( \rho \right) }{\partial \rho ^{2}}+\frac{1}{\rho }%
\frac{\partial R(\rho )}{\partial \rho }-\frac{4c^{2}}{\rho C_{2}}R(\rho )=0
\end{equation}%
whose solution is given in terms of first kind modified Bessel functions as 
\begin{equation}
R(\rho )=a_{1}I_{0}(4c\sqrt{\frac{\rho }{C}})+a_{2}K_{0}(4c\sqrt{\frac{\rho 
}{C}}),
\end{equation}%
in which $a_{1}$ and $a_{2}$ are the integration constants. The square
integrability of the solution (51) is checked by calculating the squared
norm defined in Eq. (45). In the limiting case of $\rho \rightarrow 0,$ the
squared norm reduces to the following form 
\begin{equation}
\Vert R(\rho )\Vert ^{2}\sim \int_{0}^{const.}\rho \left\vert R(\rho
)\right\vert ^{2}d\rho .
\end{equation}%
Calculations have revealed that the squared norm $\Vert R(\rho )\Vert
^{2}<\infty ,$ indicating that the solution is square integrable and belongs
to the Hilbert space.

\subsubsection{The case when $\protect\rho \rightarrow \infty $}

In the case when $\rho \rightarrow \infty ,$ the Eq. (49) simplifies to,

\begin{equation}
\frac{\partial ^{2}R\left( \rho \right) }{\partial \rho ^{2}}+\frac{2}{\rho }%
\frac{\partial R(\rho )}{\partial \rho }+\frac{\alpha }{\rho ^{2}}R(\rho )=0,
\end{equation}%
whose solution is%
\begin{equation}
R(\rho )=\frac{a_{3}\rho ^{\frac{\gamma }{2}}+a_{4}\rho ^{-\frac{\gamma }{2}}%
}{\sqrt{\rho }},\text{\ \ \ \ \ \ \ }
\end{equation}%
in which $\gamma =\sqrt{1-4\alpha }$ , and $a_{3}$ and $a_{4}$ are the
integration constants.

The square integrability of the solution (54) is checked by calculating the
squared norm defined in equation (45) in the limiting case of the metric
(10) when $\rho \rightarrow \infty ,$ which is given by%
\begin{equation}
\Vert R\left( \rho \right) \Vert ^{2}\sim \int_{const.}^{\infty }\left\vert
R\left( \rho \right) \right\vert ^{2}d\rho .
\end{equation}

The analysis has indicated that depending on the values of the complex
constant $\alpha $ together with the integration constants $a_{3}$ and $%
a_{4} $, there are some specific solutions that fails to satisfy square
integrability condition (i.e., $\Vert R\left( \rho \right) \Vert
^{2}\rightarrow \infty $), However, for a generic case the solution is
square integrable and belongs to the Hilbert space.

The method of defining whether the operator $A$ has a unique self-adjoint
extension (or essentially self-adjoint) or not is to investigate the
solution of Eq. (49) in the entire space $\left( 0,\infty \right) $ and
count the number of solutions that do not belong to the Hilbert space. In
other words, if there is one solution that fails to be square integrable for
the entire space then the operator $A$ is said to be essentially
self-adjoint. Our analysis has shown that although the behaviour of Eq.
(49), when $r\rightarrow \infty ,$ admits solution that is not square
integrable, generically, it is square integrable. Hence, the operator $A$ is
not essentially self-adjoint and the future time evolution of the quantum
particles/waves can not be predicted uniquely. Consequently, the classical
naked singularity persists and remains quantum mechanically singular when
probed with massive bosons described by the Klein-Gordon equation.

\subsection{Dirac Fields}

The timelike naked singularity of the considered spacetime is probed also
with spinorial fields ($spin-1/2$) obeying the Dirac equation. The adopted
method for the solution of the Dirac equation in $2+1-$dimensional curved
geometry was given in \cite{23}. This method has been successfully used in
earlier studies along this line \cite{23,27}. The Dirac equation in $2+1$
dimensional curved geometry for a free particle with mass $m$ is given by,

\begin{equation}
i\sigma ^{\mu }\left( x\right) \left[ \partial _{\mu }-\Gamma _{\mu }\left(
x\right) \right] \Psi \left( x\right) =m\Psi \left( x\right) ,
\end{equation}%
where $\Gamma _{\mu }\left( x\right) $\ is the spinorial affine connection
given by

\begin{equation}
\Gamma _{\mu }\left( x\right) =\frac{1}{4}g_{\lambda \alpha }\left[ e_{\nu
,\mu }^{\left( i\right) }(x)e_{\left( i\right) }^{\alpha }(x)-\Gamma _{\nu
\mu }^{\alpha }\left( x\right) \right] s^{\lambda \nu }(x),
\end{equation}%
with

\begin{equation}
s^{\lambda \nu }(x)=\frac{1}{2}\left[ \sigma ^{\lambda }\left( x\right)
,\sigma ^{\nu }\left( x\right) \right] .
\end{equation}

Since the fermions have only one spin polarization in $2+1$-dimension, the
Dirac matrices $\gamma ^{\left( j\right) }$ can be expressed in terms of
Pauli spin matrices $\sigma ^{\left( i\right) }$ so that

\begin{equation}
\gamma ^{\left( j\right) }=\left( \sigma ^{\left( 3\right) },i\sigma
^{\left( 1\right) },i\sigma ^{\left( 2\right) }\right) ,
\end{equation}%
where the Latin indices represent internal (local) frame. In this way,

\begin{equation}
\left\{ \gamma ^{\left( i\right) },\gamma ^{\left( j\right) }\right\} =2\eta
^{\left( ij\right) }I_{2\times 2},
\end{equation}%
where $\eta ^{\left( ij\right) }$\ is the Minkowski metric in $2+1$%
-dimension and $I_{2\times 2}$\ is the identity matrix. The coordinate
dependent metric tensor $g_{\mu \nu }\left( x\right) $\ and matrices $\sigma
^{\mu }\left( x\right) $\ are related to the triads $e_{\mu }^{\left(
i\right) }\left( x\right) $\ by

\begin{align}
g_{\mu \nu }\left( x\right) & =e_{\mu }^{\left( i\right) }\left( x\right)
e_{\nu }^{\left( j\right) }\left( x\right) \eta _{\left( ij\right) }, \\
\sigma ^{\mu }\left( x\right) & =e_{\left( i\right) }^{\mu }\gamma ^{\left(
i\right) },  \notag
\end{align}%
where $\mu $\ and $\nu $\ stand for the external (global) indices. The
suitable triads for the metric (10) is given by,

\begin{equation}
e_{\mu }^{\left( i\right) }\left( t,\rho ,\theta \right) =diag\left( \sqrt{%
\rho \left( C-4V_{0}\rho \right) },2\sqrt{\frac{\rho }{C-4V_{0}\rho }},\rho
\right) ,
\end{equation}%
The coordinate dependent gamma matrices and the spinorial affine connection
are given by

\begin{align}
\sigma ^{\mu }\left( x\right) & =\left( \frac{\sigma ^{\left( 3\right) }}{%
\sqrt{\rho \left( C-4V_{0}\rho \right) }},\frac{i\sigma ^{\left( 1\right) }}{%
2}\sqrt{\frac{C-4V_{0}\rho }{\rho }},\frac{i\sigma ^{\left( 2\right) }}{\rho 
}\right) , \\
\Gamma _{\mu }\left( x\right) & =\left( -\frac{\left( C-8V_{0}\rho \right)
\sigma ^{\left( 2\right) }}{8\rho },0,\frac{i\sigma ^{\left( 3\right) }}{4}%
\sqrt{\frac{C-4V_{0}\rho }{\rho }}\right) .  \notag
\end{align}%
Now, for the spinor

\begin{equation}
\Psi =\left( 
\begin{array}{c}
\psi _{1} \\ 
\psi _{2}%
\end{array}%
\right) ,
\end{equation}%
the Dirac equation can be written as%
\begin{equation}
\frac{i}{\sqrt{\rho \Sigma \left( \rho \right) }}\frac{\partial \psi _{1}}{%
\partial t}-\sqrt{\frac{\Sigma \left( \rho \right) }{4\rho }}\frac{\partial
\psi _{2}}{\partial \rho }+\frac{i}{\rho }\frac{\partial \psi _{2}}{\partial
\theta }+\left( \frac{C-8V_{0}\rho }{8\rho \sqrt{\rho \Sigma \left( \rho
\right) }}-\sqrt{\frac{\Sigma \left( \rho \right) }{16\rho ^{3}}}\right)
\psi _{2}=m\psi _{1}
\end{equation}%
and%
\begin{equation}
-\frac{i}{\sqrt{\rho \Sigma \left( \rho \right) }}\frac{\partial \psi _{2}}{%
\partial t}-\sqrt{\frac{\Sigma \left( \rho \right) }{4\rho }}\frac{\partial
\psi _{1}}{\partial \rho }-\frac{i}{\rho }\frac{\partial \psi _{1}}{\partial
\theta }+\left( \frac{C-8V_{0}\rho }{8\rho \sqrt{\rho \Sigma \left( \rho
\right) }}-\sqrt{\frac{\Sigma \left( \rho \right) }{16\rho ^{3}}}\right)
\psi _{1}=m\psi _{2}
\end{equation}%
in which $\Sigma \left( \rho \right) =C-4V_{0}\rho .$ The coupled equations
(65) and (66), can be solved by employing the following ansatz for the
positive frequency solutions which was also used in the studies \cite{23,27}

\begin{equation}
\Psi _{n,E}\left( t,x\right) =\left( 
\begin{array}{c}
R_{1n}(\rho ) \\ 
R_{2n}(\rho )e^{i\theta }%
\end{array}%
\right) e^{in\theta }e^{-iEt}.
\end{equation}%
In terms of radial function, equations (65) and (66) transforms into,%
\begin{equation}
R_{1n}^{\prime }(\rho )+\mathcal{A}\left( \rho \right) R_{1n}\left( \rho
\right) +\mathcal{B}\left( \rho \right) R_{2n}(\rho )e^{i\theta }=0
\end{equation}%
and%
\begin{equation}
R_{2n}^{\prime }(\rho )+\mathcal{C}\left( \rho \right) R_{2n}(\rho )+%
\mathcal{D}\left( \rho \right) R_{1n}(\rho )e^{-i\theta }=0.
\end{equation}%
Equations (68) and (69), can be cast into the following decoupled equations,%
\begin{equation}
R_{1n}^{\prime \prime }(\rho )+\left\{ \mathcal{A}\left( \rho \right) +%
\mathcal{C}\left( \rho \right) -\frac{\mathcal{B}^{\prime }\left( \rho
\right) }{\mathcal{B}\left( \rho \right) }\right\} R_{1n}^{\prime }(\rho
)+\left\{ \mathcal{A}\left( \rho \right) \mathcal{C}\left( \rho \right) -%
\mathcal{B}\left( \rho \right) \mathcal{D}\left( \rho \right) +\mathcal{B}%
\left( \rho \right) \left( \frac{\mathcal{A}\left( \rho \right) }{\mathcal{B}%
\left( \rho \right) }\right) ^{\prime }\right\} R_{1n}(\rho )=0
\end{equation}%
and%
\begin{equation}
R_{2n}^{\prime \prime }(\rho )+\left\{ \mathcal{A}\left( \rho \right) +%
\mathcal{C}\left( \rho \right) -\frac{\mathcal{D}^{\prime }\left( \rho
\right) }{\mathcal{D}\left( \rho \right) }\right\} R_{2n}^{\prime }(\rho
)+\left\{ \mathcal{A}\left( \rho \right) \mathcal{C}\left( \rho \right) -%
\mathcal{B}\left( \rho \right) \mathcal{D}\left( \rho \right) +\mathcal{D}%
\left( \rho \right) \left( \frac{\mathcal{C}\left( \rho \right) }{\mathcal{D}%
\left( \rho \right) }\right) ^{\prime }\right\} R_{2n}(\rho )=0
\end{equation}%
where 
\begin{eqnarray}
\mathcal{A}\left( \rho \right) &=&\frac{C}{4\rho \Sigma \left( \rho \right) }%
-\frac{2n}{\sqrt{\rho \Sigma \left( \rho \right) }},\text{ \ \ \ \ \ \ \ \ \ 
}\mathcal{B}\left( \rho \right) =\frac{2m\sqrt{\rho }}{\sqrt{\Sigma \left(
\rho \right) }}+\frac{2E}{\Sigma \left( \rho \right) }, \\
\mathcal{C}\left( \rho \right) &=&\frac{C}{4\rho \Sigma \left( \rho \right) }%
-\frac{2\left( n+1\right) }{\sqrt{\rho \Sigma \left( \rho \right) }},\text{
\ \ \ \ \ \ \ \ \ }\mathcal{D}\left( \rho \right) =\frac{2m\sqrt{\rho }}{%
\sqrt{\Sigma \left( \rho \right) }}-\frac{2E}{\Sigma \left( \rho \right) }. 
\notag
\end{eqnarray}%
Note that the prime denote the derivative with respect to $\rho .$ The
solutions of the radial part of the Dirac equations should be analyzed for a
unique self-adjoint extension for all space $L^{2}\left( 0,\infty \right) .$
Because of the complexity of Eqs. (70,71), the behaviour of the Dirac fields
will be investigated near $\rho \rightarrow 0$ and $\rho \rightarrow \infty
. $

\subsubsection{ The case when $\protect\rho \rightarrow 0$}

The behaviour of the radial part of the Dirac equations (70) and (71) near $%
\rho \rightarrow 0$ is given by%
\begin{equation}
R_{in}^{\prime \prime }(\rho )+\frac{1}{2\rho }R_{in}^{\prime }(\rho )-\frac{%
3}{16\rho ^{2}}R_{in}(\rho )=0\text{ \ \ \ \ \ \ \ where \ \ \ \ \ \ }i=1,2%
\text{\ \ \ }
\end{equation}%
whose solutions are%
\begin{equation}
R_{in}(\rho )=a_{5}r^{(3/4)}+a_{6}r^{(1/4)}\text{ \ where \ \ \ \ \ \ }i=1,2
\end{equation}%
The square integrability is checked with the following norm written for the
case of $\rho \rightarrow 0,$ 
\begin{equation}
\Vert R_{in}(\rho )\Vert ^{2}\sim \int_{0}^{const.}\rho \left\vert
R_{in}(\rho )\right\vert ^{2}d\rho \text{ \ \ \ \ \ \ \ \ \ where \ \ \ \ \
\ }i=1,2
\end{equation}

The analysis has revealed that the squared norm is finite ( i.e. $\Vert
R\left( \rho \right) \Vert ^{2}<\infty $ ) indicating that the solution is
belonging to the Hilbert space.

\subsubsection{The case when $\protect\rho \rightarrow \infty $}

The behaviour of the radial part of the Dirac equations (73) and (74) near $%
\rho \rightarrow \infty $ is given by%
\begin{equation}
R_{in}^{\prime \prime }(\rho )-\frac{1}{\rho \sqrt{\left\vert
V_{0}\right\vert }}R_{in}^{\prime }(\rho )-\frac{2m^{2}}{\left\vert
V_{0}\right\vert }R_{in}(\rho )=0\text{ \ \ \ \ \ \ \ where \ \ \ \ \ \ }%
i=1,2\text{\ \ \ }
\end{equation}%
whose solutions are%
\begin{equation}
R_{in}(\rho )=a_{7}r^{\kappa }I_{\kappa }\left( \frac{m\sqrt{2}}{\sqrt{%
\left\vert V_{0}\right\vert }}r\right) +a_{8}r^{\kappa }K_{\kappa }\left( 
\frac{m\sqrt{2}}{\sqrt{\left\vert V_{0}\right\vert }}r\right) \text{ \ \
where \ \ \ \ \ \ }i=1,2
\end{equation}%
in which $\kappa =\frac{1+\sqrt{\left\vert V_{0}\right\vert }}{2\sqrt{%
\left\vert V_{0}\right\vert }},$ $a_{5}...a_{8}$ are the integration
constants. The square integrability is checked with the following norm
written for the case of $\rho \rightarrow \infty ,$ 
\begin{equation}
\Vert R_{in}(\rho )\Vert ^{2}\sim \int_{const.}^{\infty }\left\vert
R_{in}(\rho )\right\vert ^{2}d\rho \text{ \ \ \ \ \ \ \ \ \ where \ \ \ \ \
\ }i=1,2
\end{equation}%
The analysis has shown that the solution fails to satisfy square
integrability condition if one choose $a_{7}\neq 0$ but $a_{8}=0.$ On the
other hand, the solution is square integrable if we take $a_{7}=0$ but $%
a_{8}\neq 0.$ The generic conclusion from this analysis is that the
classical naked singularity remains quantum mechanically singular when
probed with spinor fields obeying Dirac equations.

\section{Conclusion and Discussion}

In this paper, we analyzed in details, the recently obtained solution in $%
2+1-$ dimensional gravity that describes a radial solution coupled with a
real scalar field \cite{10}. The notable feature of this solution is that
depending on the values of the parameters, one may get black hole or naked
singular solutions. The black hole properties are analyzed classically by
calculating null and null circular geodesics. On the other hand, quantum
analysis is carried out by probing the timelike naked singularity that forms
for particular values of the parameters with quantum particles. We
investigate the geodesic motion of a massless particle in the gravitational
field of a black hole. The results reveal the typical behaviour of null
geodesics around a black hole. The significance of the naked singularity is
analyzed by sending quantum particles to the timelike naked singularity that
obeys the Klein-Gordon and Dirac equations. Our analysis have indicated
that, in the generic case, the singularity is persistent and remains quantum
mechanically singular with respect to the quantum particle probes. In our
earlier study \cite{27}, we investigated the quantum singularity structure
of the SS solution. In that study, we showed that the classical singularity
remains singular against bosonic/scalar probe, but it becomes quantum
regular with respect to the fermionic probe. As stated in Section II, the
solution analyzed in this article is a one-parameter generalization of the
SS solution and this parameter has an influence on the curvature of the
resulting spacetime. The square of Ricci tensor in the SS solution was given
in \cite{8}, \ as $R_{\mu \nu }R^{\mu \nu }=\frac{2}{r^{4}}$ , whereas in
the considered solution it is calculated and given by $R_{\mu \nu }R^{\mu
\nu }=\frac{2V_{0}^{2}}{r^{4}}+\frac{C^{2}}{r^{6}}.$ The presence of the
parameter $C$ increases the rate of divergence of Ricci tensor. This means
that the singularity in the considered solution \cite{10} is stronger than
the singularity of the SS solution. As a consequence, fermionic/scalar probe
of the singularity does not help to heal the singularity and remains quantum
singular.

\end{document}